\documentclass[10pt]{iopart}
\usepackage{iopams}
\usepackage{amstext}

\usepackage{xcolor}
\usepackage{graphicx}
\begin{document}

\title{Source of pure proton beams}

\author{S.V. Golubev$^{a)}$, I.V. Izotov$^{a), b)}$, V.A. Skalyga$^{a), b)}$, S.S. Vybin$^{a), b)}$, E.M. Kiseleva$^{a), b)}$, R.L. Lapin$^{a)}$, S.V. Razin$^{a)}$, A.F. Bokhanov$^{a)}$, M.Yu. Kazakov$^{a)}$, S.P. Shlepnev$^{a)}$}
\address{$^{a)}$Federal Research Center Institute of Applied Physics of the Russian Academy of Sciences, \\ Nizhny Novgorod, Ulyanova St., 46, Russia}
\address{$^{b)}$Lobachevsky State University of Nizhny Novgorod - National Research University (UNN), \\ Nizhny Novgorod, Gagarina Av., 23, Russia}
\ead{}
\vspace{10pt}
\begin{indented}
\item[]June 2022
\end{indented}

\begin{abstract}
In the quasi-gasdynamic high-current ion source described in this work, the plasma is sustained by high-power millimeter-wave radiation under the electron cyclotron resonance (ECR) condition. In such facilities, it is possible to achieve high volumetric energy input of up to $250$ $W/cm^3$ and obtain pure proton beams with a minimum amount of impurities and molecular ions. Experiments conducted on the GISMO facility demonstrated the possibility of a proton beam formation with a current of $50$ mA and an extremely high ($99.9$\%) content of atomic ions.
\end{abstract}

%
%
%
%
\ioptwocol

\section{Introduction}
Development of a high-current proton or deuteron beam source with a low emittance is of great necessity for the implementation in various projects in accelerator centers \cite{IFMIF, FRANZ}. Particularly high requirements are imposed on the parameters of the ion beam for the accelerator-based neutron sources. In such facilities, the beam current (the required level is hundreds of mA) determines the neutron yield, and the loss of particles from the accelerated beam, which is determined by its quality, can lead to the destruction of the accelerating structures. It is worth noting that such problems are typical for the whole variety of neutron sources using accelerated ion beams, from compact neutron generators with a beam energy of the order of 100 keV \cite{BNCT2014, DD_gen} to spallation sources with energies about 1 GeV \cite{isis, lansce}.

Traditionally, several types of ion sources are used to inject proton beams into accelerators, differing both in the method of creating plasma (arc \cite{duoplasmatron}, ECR \cite{SMIS_proton}, Penning \cite{Penning}, etc.) and in the parameters of the produced beams. A detailed description the variety of ion sources and their application is well described in a number of articles and books, see, for example, \cite{IoSo_overview}. 

One of the promising sources of protons is an ECR source with a quasi-gasdynamic regime of plasma confinement \cite{GasDyn2, GasDyn}. In such a facility, it is possible to maintain plasma with unique parameters: a density of more than $10^{13} \; \textrm{cm}^{-3}$ (which is more than an order of magnitude higher than the plasma density in traditional sources) and an electron temperature of $50-100$ eV, which is an optimal value for hydrogen dissociation and ionization. This combination of parameters is achieved by using the millimeter-wave radiation of modern gyrotrons with a power of $10-100$ kW to heat the plasma confined in compact (with a typical size of $\approx 10$ cm) open magnetic traps \cite{SMIS, GISMO}.

Earlier, the course of experiments was conducted in pulsed mode on the Simple Mirror Ion Source $37$ GHz (SMIS 37) facility. It uses the $37$ GHz gyrotron radiation with a power of $100$ kW, a pulse duration of $1$ ms and a pulse repetition rate of $0.1$ Hz. As a result, the possibility of generating hydrogen ion beams with a current of $100-500$ mA, a normalized root-mean-square emittance of $0.1$ $\pi\cdot mm \cdot mrad$, and an impurity fraction of molecular hydrogen ions of less than $6\%$ was demonstrated \cite{SMIS_proton, SMIS_MCI}. The parameters of obtained beams meet practically all the requirements for modern accelerators, except for the pulse repetition rate. In most cases, either a high pulse repetition rate at the level of tens and hundreds of Hertz or a continuous mode is required. To overcome these limitations, the Gasdynamic Ion Source for Multipurpose Operation (GISMO) facility has been created. It uses the radiation of a gyrotron with a frequency of $28$ GHz and a power of up to $10$ kW operating in a continuous mode \cite{GISMO, GISMO_emit}. The main difference between the GISMO and conventional continuous ECR ion sources is associated with the possibility of implementing a discharge in a quasi-gasdynamic confinement regime and a high volumetric energy input. This peculiarity allows obtaining ion beams with the intensity that corresponds the requirements for modern applications. Moreover, it ensures efficient cleaning of the discharge chamber, which opens up the possibility of extracting pure hydrogen ion beams with an extremely high content of protons and a minimum amount of impurities and molecular ions.

The development of methods for generating pure proton beams makes it possible to directly inject the beam into the accelerator, avoiding the use of separation systems. Such systems detach the beam from molecular and impurity ions immediately before injection into the accelerator sections. Thus, the use of pure proton beams allows elimination of bending magnet which may be a noticeable source of the beam losses and degrades the beam quality.
The presented work is devoted to the study of the possibilities of realizing such a "clean" source with protons operating in a continuous mode.

\section{Experimental scheme, methods, and approaches}

\begin{figure*}\centering
	\vspace{0cm} 
    \includegraphics[width=150mm]{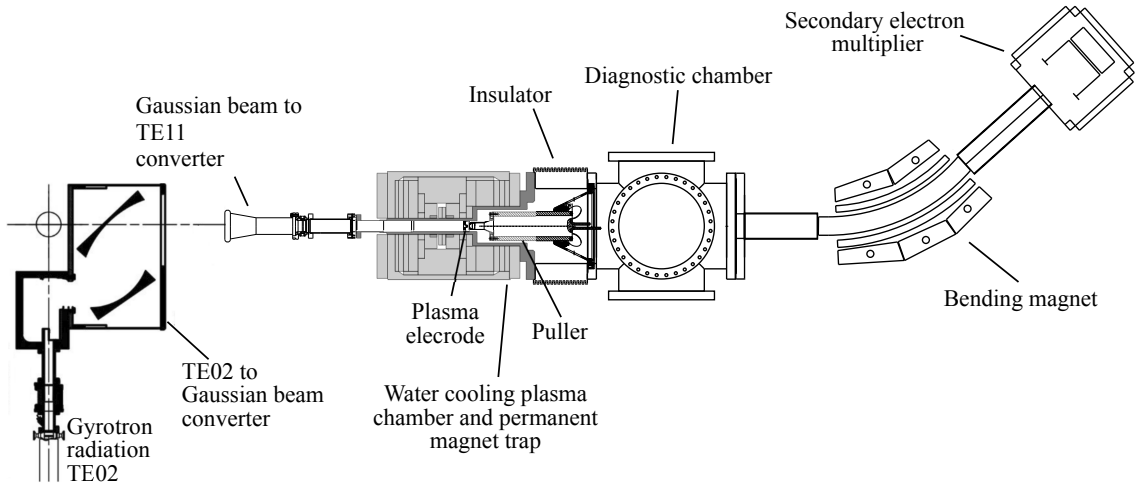}
	\caption{\label{fig1} The experimental scheme.}
	\vspace{0cm} 
\end{figure*}

The experiments were carried out on the GISMO facility, the scheme of which is shown in Fig. \ref{fig1}. A continuous technological gyrotron \cite{Gyrotron1, Gyrotron2, Gyrotron3} with a radiation frequency of $28$ GHz and a power of up to $10$ kW was used to heat the plasma. The transition to slightly lower radiation frequencies compared to the earlier experiments \cite{SMIS_proton, SMIS, SMIS_MCI, SMIS75} made it possible to switch from a superconducting magnet to a ``warm'' oil-cooled one in the gyrotron system.
\begin{figure*}\centering
	\vspace{0cm} 
    \includegraphics[width=130mm]{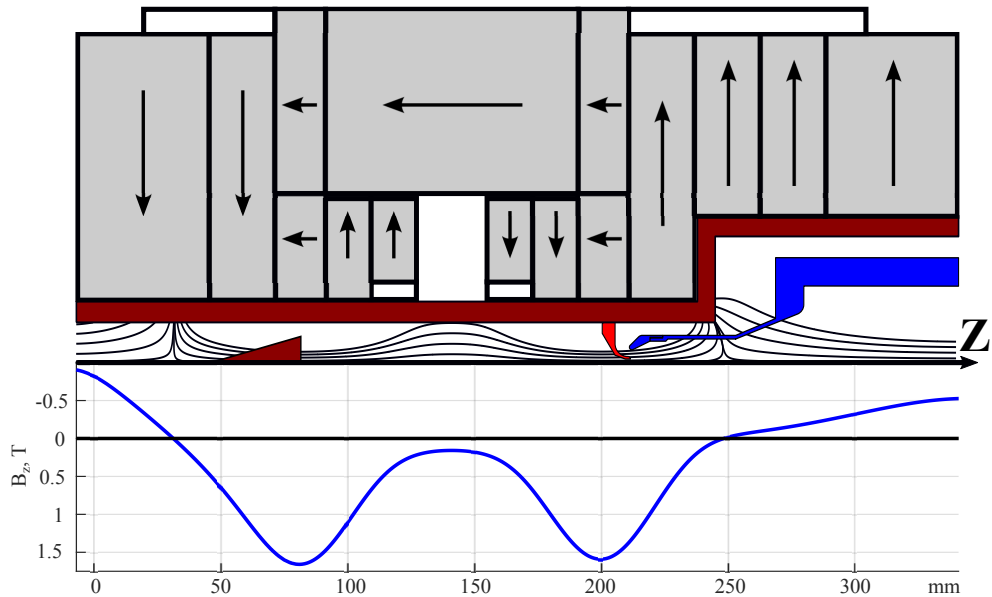}
	\caption{\label{fig2} Distribution of the magnetic field along the axis of the system. Black contours are the boundaries of magnetic blocks.}
	\vspace{0cm} 
\end{figure*}
To confine the plasma, a magnetic trap based on permanent magnets was developed, which made it possible to implement a continuous mode of operation of the ion source or a mode with a high pulse repetition rate. The configuration of the magnetic field in the plasma confinement region is close to that of a simple magnetic trap with a length of $12$ cm and a mirror ratio of 6. The computational model of the magnetic field distribution in the system is shown in Fig. \ref{fig2} together with the contours of the magnets and the plasma vacuum chamber. The walls of the plasma chamber at the minimum of the magnetic field strength, i.e. in the center, are a natural limiter for the plasma. In this configuration, the volume of heated and confined plasma is about $40$ $cm^3$, which is tenfold lower than in conventional ECR ion sources. Due to such a small plasma volume, it is possible to realize a discharge with a volumetric energy input level of up to $250 \; W/cm^3$, which is a record value for continuous ECR sources. Such values of power density ensure effective cleaning of the discharge chamber and, consequently, the production of hydrogen ion beams with high brightness and extremely high purity.

\begin{figure}\centering
	\vspace{0cm} 
    \includegraphics[width=60mm]{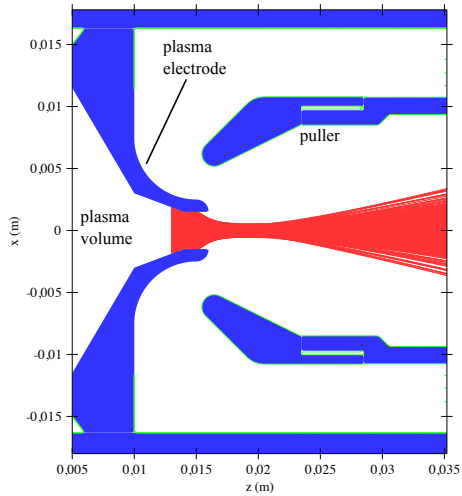}
	\caption{\label{fig3} Ion beam formation scheme.}
	\vspace{0cm} 
\end{figure}
The plasma was sustained in hydrogen at low (on the order of $1$ mTorr) pressures of the injected gas in a discharge chamber. It was constructed as a cylindrical resonator placed in the center of the magnetic trap. In the injection mirror, an electrodynamic system for microwave launching was installed. It provided the matching between the waveguide and the plasma-filled resonator and prevented the plasma flow into the waveguide. In the extraction mirror, the plasma electrode was installed with a supercritical with respect to the heating wavelength orifice. The plasma heating was realised due to the absorption of microwave radiation under conditions of electron cyclotron resonance at surfaces with $B=1$ T.

The formation of ion beams was carried out using a two-electrode (diode) extraction system with an inhomogeneous electric field~\cite{vybin2020, vybin2021}. The key feature of the extraction system is the plasma electrode which has a narrow tip. The accelerating field is higher near the plasma meniscus when compared to the traditional flat geometry, which allows for use of a lower accelerating voltage. The plasma electrode was electrically connected to the plasma chamber and was at a high potential (at a level of $20-30$ kV), while the puller electrode was grounded. The experiments were carried out in a pulsed extraction mode with a high voltage pulse duration of $20$ ms. A detailed diagram of the beam formation system is shown in Fig. \ref{fig3}.

The diameter of the plasma electrode hole is $3$ mm, the diameter of the puller hole is $10$ mm, and the distance between the electrodes is $2.5$ mm. Total beam current was measured with a Faraday cup placed on the axis of the system inside the diagnostic chamber. The ion composition of the extracted beam was studied using a traditional method with a magneto-static analyzer. It consisted of an ion transmission line, a bending magnet, and an ion beam collector (a Faraday cup).

\section{Experimental results}
In the course of the experiments, the conditions for maintaining the discharge were varied, namely, the hydrogen pressure in the plasma chamber and the radiation power of the gyrotron. 

\begin{figure}\centering
	\vspace{0cm} 
    \includegraphics[width=80mm]{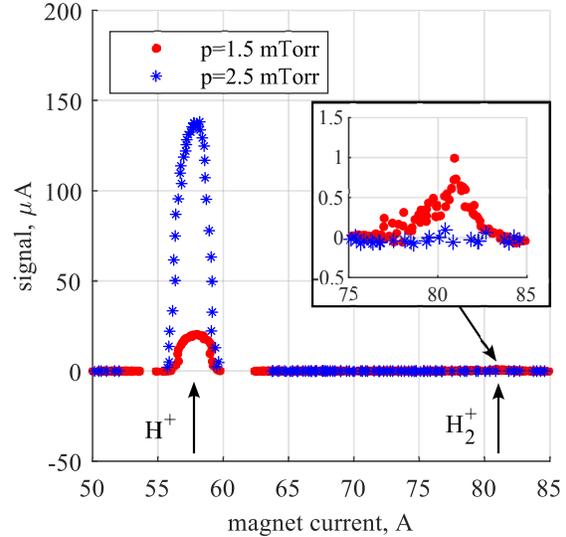}

	\caption{\label{fig4} The dependence of the ion beam current on the magnetic analyzer on the current of the magnet for two values of pressure ($1.5$ mTorr, $2.5$ mTorr). The arrows show the current values corresponding to the type of an ion that hit the Faraday cup.}
	\vspace{0cm} 
\end{figure}

Direct proofs of the possibility of obtaining pure proton beams are the results of studies of the composition of the ion beam, which are presented in Fig. \ref{fig4} for the optimal experimental conditions ($4.6$ kW of the gyrotron power).

\begin{figure}\centering
	\vspace{0cm} 
    \includegraphics[width=60mm]{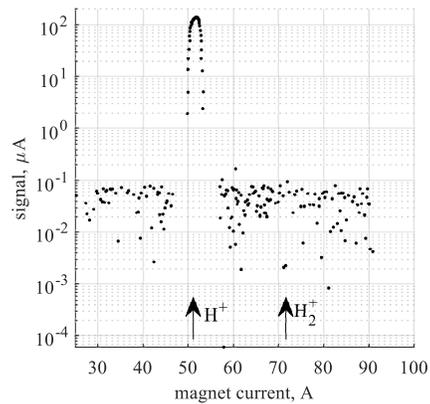}
	\caption{\label{fig5} The dependence of the ion beam current on the magnetic analyzer on the current of the magnet on a logarithmic scale.}
	\vspace{0cm} 
\end{figure}

\begin{figure}\centering
	\vspace{0cm} 
    \includegraphics[width=60mm]{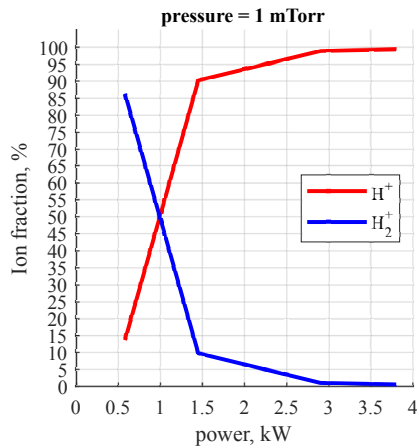}
	\caption{\label{fig6} Dependence of the ion fractions in the beam on the microwave radiation power at a fixed pressure of $1$ mTorr.}
	\vspace{0cm} 
\end{figure}

\begin{figure}\centering
	\vspace{0cm} 
    \includegraphics[width=60mm]{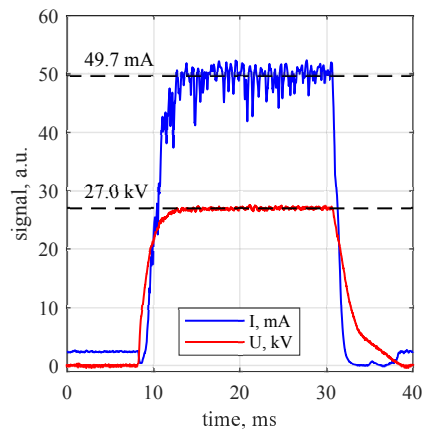}
	\caption{\label{fig7} A typical waveform observed on the oscilloscope in the experiments for the record values of the ion beam current. Ion current is marked in blue, extraction voltage is marked in red. \textcolor{blue}{The extracted beam current density is above $700~\text{mA}~\text{cm}^{-2}$.}}
	\vspace{0cm} 
\end{figure}

It can be seen that, at a lower pressure of the injected gas, along with the proton beam, molecular hydrogen ions are also observed (it is of note that no other ions were detected in the beam). With an increase in n the discharge pressure, the ion beam current increases significantly, and traces of molecular ions disappear.
Figure \ref{fig5} shows the distribution of ions over charges under optimal conditions ($4.6$ kW of the gyrotron power, $2.5$ mTorr of the gas pressure) on a logarithmic scale, which makes it possible to estimate the purity of the beam; according to the data obtained, the proportion of impurities in the proton beam does not exceed $0.1\%$.
At the characteristic pressure of the injected gas ($1$ mTorr) and with an increase in the power of microwave radiation, a decrease in the fraction of molecular ions in the beam is observed (see Fig. \ref{fig6}).
In the optimal mode, the beam current reached the value of $50$ mA which corresponds to the beam current density above $700~\text{mA}~\text{cm}^{-2}$ (at the plasma electrode). A typical waveform is shown in Fig. \ref{fig7}.
\section{Conclusion}

The experimental results presented in this work demonstrated the possibility of a proton beam formation with a current at the level of $50$ mA and an extremely high content of atomic ions (more than $99.9$\%). Ranges of parameters (namely, the power of the radiation supporting the discharge and the pressure of the injected gas), where such a regime is achieved, were determined.

\section{Acknowledgements}
This research was supported by the grant of Russian Science Foundation (project number 21-19-00844).

\section*{References}
\bibliography{references}

\end{document}